\documentclass[11pt, letter]{article}

  \usepackage{graphicx, subfigure}
  \usepackage{epstopdf}
  \usepackage{color}
  \usepackage{epsfig,amssymb,amsmath}

\usepackage{amsmath,amssymb,amsbsy,amstext,mathrsfs}
\usepackage{caption}

\definecolor{AV}{rgb}{0.65,0.0,0}
\definecolor{DT}{rgb}{0,0,0.65}
\definecolor{SB}{rgb}{0,0.65,0}

\usepackage[colorlinks=true,      linkcolor=blue,      urlcolor=blue,      filecolor=blue,      citecolor=blue,
      pdfstartview=FitH,     pdfpagemode=UseNone,      bookmarksopen=true]{hyperref}  

\usepackage[all]{hypcap}     

\usepackage{latexsym,amssymb,amsfonts,epsfig,graphicx,color}
\usepackage{bm,exscale, amsmath, mathrsfs}
\usepackage{amsfonts, amssymb}
\newcommand{\be}{\begin{equation}}
\newcommand{\ee}{\end{equation}}

\def\be{\begin{equation}}
\def\ee{\end{equation}}

\def\M{{\mathcal M}}

\def\be{\begin{equation}}
\def\ee{\end{equation}}

\def\a{\alpha}

\def\l{\lambda}
\def\k{\kappa}\def\s{\sigma}\def\l{\lambda}

\def\bg{\bar{g}}
\def\beq{\begin{eqnarray}}\def\eeq{\end{eqnarray}}
\def\ba#1\ea{\begin{align}#1\end{align}}
\def\bg#1\eg{\begin{gather}#1\end{gather}}
\def\bm#1\em{\begin{multline}#1\end{multline}}
\def\bmd#1\emd{\begin{multlined}#1\end{multlined}}
\def\bea{\begin{eqnarray}}
\def\eea{\end{eqnarray}}

\def\a{\alpha}

\def\k{\kappa}
\def\l{\lambda}

\def\s{\sigma}

\def\la{\label}

\def\fr{\frac}

\def\nn{\nonumber}

\date{}

\makeatletter

\@addtoreset{equation}{section}
\makeatother

\begin{document}

\thispagestyle{empty}

\begin{center}
{\Large \bf Internal Structure of Charged AdS Black Holes}\\[10mm]

\vspace{8mm}
\large
{\large  Srijit Bhattacharjee$^{\, a}$, Sudipta Sarkar$^{\, b}$, and  Amitabh Virmani$^{\, a}$}

\normalsize
\vspace{5mm}
{$^{a\,}$Institute of Physics\\
Sachivalaya Marg, Bhubaneshwar, Odisha, India 751005 \\}
\vspace{5mm}
{$^{b \,}$Indian Institute of Technology\\
Gandhinagar, Gujarat, India 382355 }

\vspace{10mm}
\hrule
\vspace{10mm}

\begin{tabular}{p{12cm}}
{\small
When an electrically charged black hole is perturbed its inner horizon becomes a singularity, often referred to as the Poisson-Israel mass inflation singularity. 
Ori constructed a model of this phenomenon for asymptotically flat black holes, in which the metric can be determined explicitly in the mass inflation region. 
In this paper we implement the Ori model for charged AdS black holes. 
We find that the mass function inflates faster than the flat space case as the inner horizon is approached.  
Nevertheless, the mass inflation singularity is still a weak singularity: although spacetime curvature becomes infinite, tidal distortions remain finite on physical objects attempting to cross it. 
}
\end{tabular}
\vspace{10mm}
\hrule
\end{center}

\tableofcontents

\vspace{8mm}
\hrule
\newpage

\section{Introduction}
\label{sec:intro}

Classical internal structure of  a charged or rotating black hole is very different from that of the Schwarzschild black hole. The central singularity, which is typically timelike for a charged or rotating black hole, is surrounded by an inner event horizon, which is 
a surface of infinite blueshift. Linearised perturbation gets amplified by the geometry near the inner horizon \cite{Penrose:68, Matzner, Chandra:82}. It is generally believed that the inner horizon\footnote{In this paper, the phrase inner horizon refers to the ingoing Cauchy horizon. The outgoing Cauchy horizon is also singular \cite{Marolf:2011dj}, but that topic we do not address in this work.} is a singular surface --- the singularity arising from the back reaction of the blue-shifted perturbation.  This singularity drastically changes the internal structure of the black hole.

Despite decades of work, the precise nature of the expected singularity and the resulting spacetime structure in sufficiently general settings is still far from fully understood. Detailed numerical simulations \cite{Brady:1995ni, Burko:1997zy, Hod:1998gy} and rigorous proofs \cite{Dafermos:2003wr, Dafermos:2003yw,Dafermos:2004jp} have clarified the situation considerably in asymptotically flat setting,  though little is known about the internal structure of black holes in asymptotically anti-de Sitter (AdS) or other 
boundary conditions.
A good understanding of the classical internal structure of black holes is the first step in attempting to see its manifestations in AdS/CFT scenarios. In this paper we present  a study of the internal structure of charged AdS black holes.

A key milestone in discussions of internal structure of black holes  is the Poisson-Israel  mass inflation phenomenon \cite{Poisson:1990eh},  first explored for charged asymptotically flat black holes. In their pioneering work, Poisson and Israel proposed a simple model to explore the back reaction effects of the blue-shifted perturbations on Reissner-Nordstr\"om  black hole.  This perturbation can correspond to the  back-scattering of the emission carrying away non-spherical inhomogeneities from the surface of a star undergoing gravitational collapse. The decay rate of  such inhomogeneities is governed by the Price's law  \cite{Price:1971fb, Gundlach:1993tp}: the in-falling flux decays as a power law in the advanced time $v$.

Poisson and Israel modelled the ingoing perturbation in the background of the Reissner-Nordstr\"om black hole as a spherically symmetric stream of massless particles. 
The Reissner-Nordstr\"om geometry is converted into charged Vaidya spacetime with such a stream present. The singularity of this charged Vaidya spacetime was explored by Hiscock \cite{Hiscock}, 
where it is shown that the inner horizon of a charged Vaidya spacetime is singular, though the singularity is rather weak. Various curvature scalars remain finite at the inner horizon. Although tidal forces along the path of an in falling observer diverge,  tidal distortions on a physical object attempting to cross it remain finite.

Poisson and Israel argued that this situation is drastically changed if one considers in addition to the influx, an outflux of massless particles. The outflux represents back-scattering of the ingoing particles  once they are inside the outer event horizon. In the presence of this cross-flow
an exact solution of the Einstein-Maxwell equations is not known. Nevertheless, by a detailed analysis of the equations of motion, Poisson and Israel  showed that the effective mass parameter of the internal spacetime diverges as the inner horizon is approached.  In fact, the inner horizon becomes a curvature singularity, for example, the curvature scalar $R_{abcd} R^{abcd}$ blows up at the inner horizon.

The mechanism responsible for the mass inflation phenomenon is the presence of both ingoing and outgoing fluxes of radiation. 
Shortly after the Poisson-Israel analysis, Ori \cite{Ori} developed an explicit exact solution for this phenomenon. His key observation was that the mass inflation phenomenon is not sensitive to the details of the outgoing radiation, but merely to its presence\footnote{The crucial point is that the presence of the outgoing radiation results in the separation between the inner event horizon and the inner apparent horizon.}. Ori modelled the outgoing radiation as a delta function shell, thus reducing the  Poisson-Israel   model to matching two patches of charged Vaidya spacetimes along a thin-layer of outgoing null radiation. An explicit solution can then be constructed everywhere. One finds that in this model the mass-inflation singularity is a null curvature singularity. Although some of the curvature scalars
diverge,  tidal distortions on physical objects attempting to cross it remain finite.

In this paper, we analyse the phenomenon of mass inflation along the lines of the Ori model for spherical black holes in global AdS spacetime. The rest of the paper is organised as follows. In section 
\ref{sec:ori_set_up} we review basic ingredients of the Ori model. We discuss gluing of two charged Vaidya spacetimes across an outgoing null shell. The mass function of the external charged Vaidya spacetimes is an input in this model.

Unlike the asymptotically flat case, for black holes in global AdS there are no power law tails for decaying perturbation \cite{Horowitz:1999jd}.  As a first guess, one might expect that the late time decay is governed by the lowest lying quasinormal mode. With this as motivation, in section \ref{sec:mass_inflation}  we  first assume an exponential decay behaviour of the mass function with respect to  the advanced time in the external charged Vaidya spacetime and analyse the Ori model. 
We take a weak exponential fall-off for this analysis, 
as in the large angular momentum limit quasinormal modes are long lived \cite{Festuccia:2008zx, Berti:2009wx, Dias:2012tq}.

Interestingly, such a simple guess may not be the correct answer. Holzegel and Smulevici \cite{Holzegel:2011uu} showed that the decay of generic scalar perturbation in Kerr AdS spacetimes is actually logarithmic. Motivated by this analysis, in section \ref{sec:mass_inflation}  we also study the Ori model of mass inflation with logarithmic fall-off  behaviour of the mass function. Arguably, our model overlooks many aspects of the Holzegel-Smulevici analysis,  it does capture some of the key features.  We end with a brief discussion in section \ref{sec:conclusions}. In appendix \ref{app1} more details on the Ori model are presented, and in appendix \ref{app2} strength of the mass-inflation singularity is analysed. Whether our model captures the physics of mass-inflation phenomenon correctly in AdS can only be confirmed by future numerical or mathematical work. 

\section{Ori model: the set-up}
\label{sec:ori_set_up}

We start with the charged Vaidya solution in $d$ space-time dimensions,
\begin{equation}
 ds^2=-f(r,v)dv^2 +2dv dr +r^2 d\Omega_{d-2}^2.
\la{vaidya} \end{equation}
The metric function $f(r,v)$ is
\be
f(r,v)=1-\frac{2 m(v)}{r^{d-3}}+\frac{q^2}{r^{2(d-3)}}+ \frac{r^2}{l^2},
\ee
with AdS length $l^2= -\frac{(d-1)(d-2)}{2 \Lambda}$ and  $\Lambda$ is the cosmological constant. The constant $q$ parameterizes the electric charge of the black hole, and $v$ is the advanced null coordinate. The function $m(v)$ represents dependence of hole's mass on the advanced time; it is determined by the flux of ingoing radiation. We shall assume that the function $f$ always has two real positive roots in the static limit. In order for this to happen the mass, charge, and $\Lambda$ must satisfy certain constraints, a discussion of which can be found in \cite{Brecher:2004gn}.

\begin{figure}[t]
 \centering
    \includegraphics[width=0.5\textwidth]{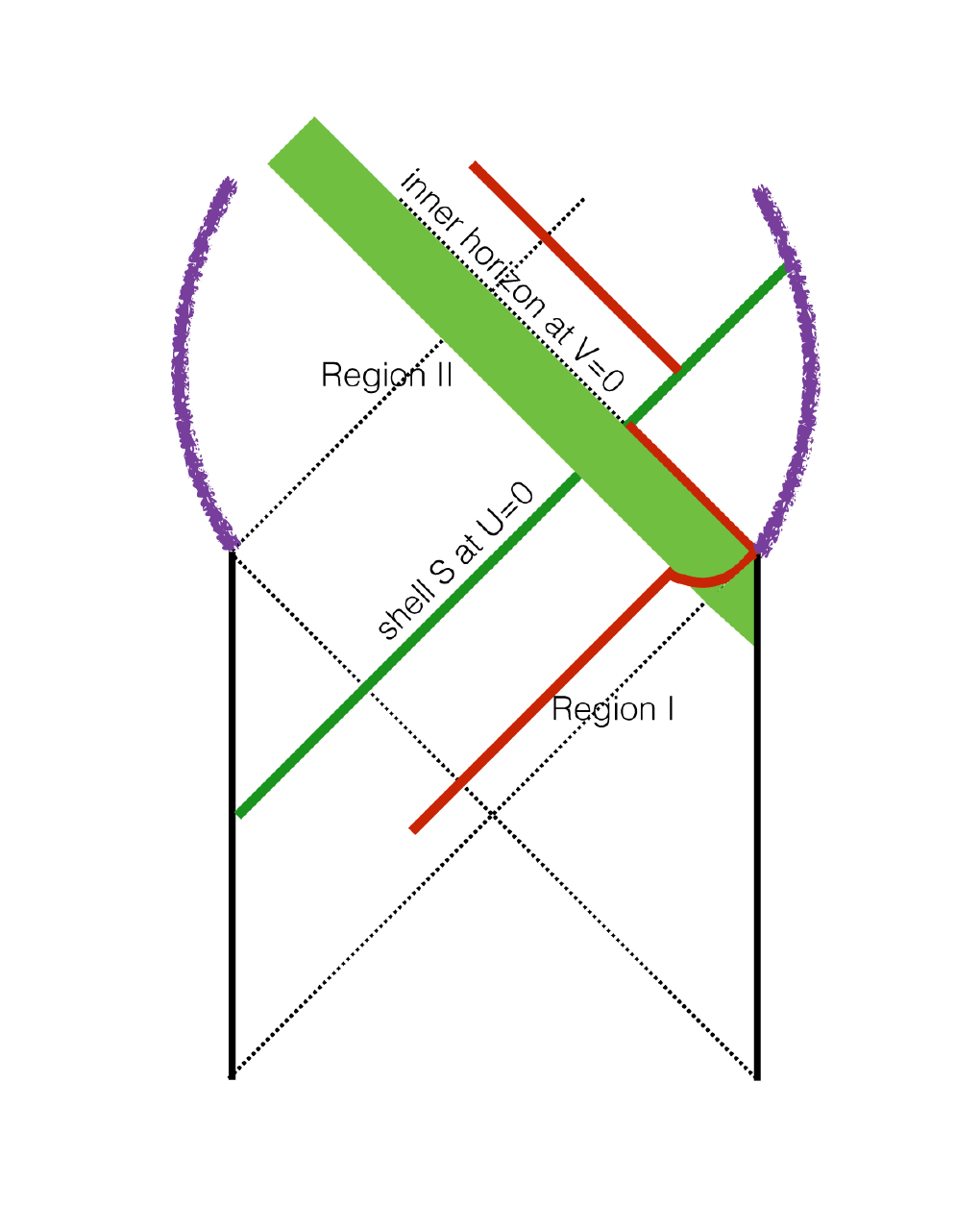}
      \caption{Penrose diagram of the spacetime formed by matching two charged Vaidya spacetimes along the thin null shell $S$ (thick green line). The thick red lines are the inner and the outer apparent horizons. The dotted lines are the event horizons of the background geometry (mass $m_f$, cf.~\eqref{mass_function}).}
    \label{figure1}
\end{figure}

The Ori model consists in two charged Vaidya solutions respectively in region I and region II matched along the null shell $S$, cf.~Fig.~\ref{figure1}. The shell $S$ is situated between the outer  and (outgoing) inner horizon. Let $v_1$ and $v_2$  be the advanced times of region I and II respectively.  The preliminary junction condition across the null shell requires that the metric tensor be continuous across $S$. Thus coordinate $r$ is continuous. Furthermore, the null fluid along the shell is assumed to be electrically neutral and pressureless. This ensures that $\lambda$ is an affine parameter of both sides of the shell \cite{Barrabes:1991ng, Poisson, Ori}.

Let us assume that the shell's null generators are parametrized by an affine parameter $\lambda$ which is zero at the inner horizon and is negative below it, i.e., $\lambda$ increases with time coordinates $v_i, \ i=1, 2$ . The null generators are characterized by $R(\lambda)
$ --- the radius of the shell $S$ --- and an advanced time $v_i(\lambda)$. The nullity condition of the shell demands,
\be
\fr {\dot{R}(\l)}{ \dot{v}_i(\l)}={1\over2} f_i(R(\l), v_i (\lambda)),
\la{nullc}
\ee
where over-dot denotes derivative with respect to  $\l$.  The geodesic equation for null generators gives,
\be
\ddot{v}_i=-{1\over2} [\dot{v}_i(\l)]^2 \left(\partial_r f_i(r,v)\right) \Big|_{r = R(\lambda), v= v_i(\lambda)} .
\la{geod}
\ee

The two equations, namely  the nullity condition \eqref{nullc} and the geodesic equation \eqref{geod}, can be combined to give equations that can be readily integrated. Defining 
\be
z_i (\lambda)=\fr{R(\l)^{d-3}}{\dot{v}_i},
\label{zdef}
\ee
the  nullity condition \eqref{nullc}  can be recast as
\be
z_i (\lambda) \dot{R}(\l)=\M(R(\l)) - m_i(v_i (\lambda)),
\ee
where
it is convenient to define a function $\M(r)$ 
\be
\M(r):= \fr{1}{2}r^{d-3}\left[1 + \fr{q^2}{r^{2(d-3)}} + \fr{r^2}{l^2}\right] = m_i(v_i) + \frac{1}{2} r^{d-3} f_i(r,v_i).
\label{nullc2}
\ee
Differentiating $z_i (\lambda)$ with respect to $\l$, the geodesic equation \eqref{geod} can be recast as,
\be
\dot{z}_i (\lambda) = \frac{1}{2} \left[\partial_r (r^{d-3} f_i(r,v))\right] \Big|_{r = R(\lambda), v= v_i(\lambda)} = \partial_r\M(r) \Big|_{r = R(\lambda)}.
\label{geod2}
\ee
Equations \eqref{zdef}, \eqref{nullc2}, and \eqref{geod2} are the key equations; they give rise to the following three matching equations:
\bea
z_i(\l)&=& Z_i\,+\, \int_0^{\l} \partial_r\M(r) \Big|_{r = R(\xi)}d\xi, \label{m1}\\
v_i(\l)&=& \int^{\l} \fr {R^{d-3}(\xi)}{z_i(\xi)}d\xi, \label{m2}\\
m_i(v_i(\l))&=&\M(R(\l))\,-\,z_i (\l) \dot{R}(\l) \label{m3}.
\eea

In equation \eqref{m1} the $Z_i$'s are integration constants. We have ignored any integration constant associated with $v$ as it is irrelevant --- adding a constant to coordinate $v_i$ does not make any physical difference. 
Let the mass functions of two patches are related by
\begin{equation}
 m_2(v_2)=m_1(v_1) + \Delta m(v_2),
\end{equation}
where $\Delta m(v_2)$ is the mass energy of outgoing null shell $S$. 
 Equation (\ref{m3}) yields, 
\be 
m_2(v_2(\lambda)) - m_1(v_1(\lambda))=\Delta m (v_2(\l))=(Z_1-Z_2)\dot{R}(\l) \la{minf}.
\ee

The constant $Z_1$ and the function $R(\l)$ are determined by the fact that in region I, charged Vaidya solution is nothing but a slightly perturbed Reissner-Nordstr\"om solution. We write the mass function in the form
\be
m_1(v_1) = m_f- \delta m_1(v_1),
\label{mass_function}
\ee
where $m_f$ is the final mass of the hole, and  $\delta m_1(v_1)$ represents the mass associated with the radiative tail which dominates the late-time behaviour of decaying perturbation. As $v_1 \to \infty$, $\delta m_1(v_1) \to 0$, i.e., $m_1(v_1) \to m_f$.

The inner (Cauchy) horizon of the background geometry  is at $r=r_-, v_1 \to \infty$. The numerical value of $r_-$ is determined by the final mass $m_f$ of the black hole.
We evaluate equations (\ref{m1}) and (\ref{m2}) in the vicinity of the inner horizon, i.e., as the affine parameter on the shell  $\l$ approaches zero. Substituting $R(\l) \approx r_-$, equation (\ref{m1}) gives 
\be
z_i(\lambda) \approx Z_i -  \kappa_- r_-^{d-3}\l,
\ee
where 
\be
\kappa_-:=-{1 \over {r_-}^{d-3}} \partial_r \M(r) \Big|_{r=r_-} = - \frac{1}{2} \partial_r f_1(r,v_1 \to \infty)\Big|_{r=r_-}.
\ee 
We also have from equation (\ref{m2}),
\be 
v_i(\l) \approx \int^\l \fr{r_-^{d-3}}{Z_i-\k_- r_-^{d-3}\xi} d\xi=-{1 \over \k_- } \ln \left| \frac{Z_i}{\k_-r_-^{d-3}} - \l \right|
\ee
 Since as $v_1 \to +\infty$, $\l\rightarrow 0$, we must have $Z_1=0$. This implies,
\be
\l \approx - e^{-\k_- v_1},
\la{lamb}
\ee
in the vicinity of the inner horizon. The relation between the advanced time $v_2$ and $\lambda$ can also be determined from equation (\ref{m2}) and using the fact that in the  region\footnote{For the shell $S$ to be non-trivial we must have $Z_2 \neq 0$, cf.~\eqref{minf}.} $\l \approx 0$, $R(\lambda) \approx r_-$,
\be 
v_2(\l)\approx \fr{r_-^{d-3} \l}{Z_2}. \label{v2_lambda}
\ee
Since $v_2$ is proportional to $\lambda$, the inner horizon in the region II is at $v_2$ equals to zero. The mass inflation corresponds to the statement that $m_2(v_2)$ diverges as$v_2$ approaches zero.

Equation \eqref{minf} now becomes
\be 
\Delta m (v_2(\l))=-Z_2\dot{R}(\l) \la{minf2}.
\ee
Since $\dot{R}(\l)$ is negative inside the hole, we must have $Z_2$ to be positive in order to have positive mass energy for the thin-shell. To determine $R(\l)$ we recall 
from the nullity condition \eqref{nullc} that,
\be
\fr {dR(v_1)}{ dv_1}={1\over2} f_1(R(v_1),v_1), 
\la{Req1}
\ee
or equivalently, 
\be
\fr {dR(v_1)}{ dv_1}=\fr{1}{ R(v_1)^{d-3}}\left[\M(R(v_1))-m_1(v_1)\right]
.\la{Req}
\ee
The late-time solution of this equation clearly depends on the nature of the mass function \eqref{mass_function}. We address the issue of the choice of the mass function and the integration of the radial equation in the next section.  Expanding $\M(R(v_1))$ in equation (\ref{Req}) around $r=r_-$ with 
\be
R(v_1)=r_- + \delta R(v_1),
\ee 
we get,
\be
\fr{d}{d v_1} \delta R(v_1) \approx {1\over r
_-^{d-3}}\left[-\k_- r_-^{d-3}\delta R(v_1) + \delta m_1(v_1)\right]. \la{linr}
\ee
In obtaining this equation we have used the fact that $m_1(v_1)+\delta m_1(v_1)=\M(r_-)=m_f$, the final mass of the hole.

Equation \eqref{linr} is the final equation for this section. So far we have not used any information about the asymptotic conditions of the spacetime. Therefore, this analysis is sufficiently general and is applicable to AdS black holes as well. In the next section we  consider  AdS boundary conditions and evaluate the growth of the perturbation near the Cauchy horizon.

\section{Mass Inflation for  Charged AdS Black Hole}
\label{sec:mass_inflation}

The late time behaviour of a decaying field in the charged AdS black hole background is the key ingredient needed to proceed further. In asymptotically flat case,  Price's law dictates the late time behaviour of a decaying field. The specific power law decay behaviour arises due to scattering off a weak Coulomb potential near infinity. For asymptotically AdS black hole, the potential diverges at infinity.  Reference \cite{Ching:1995tj}  studied the late time behaviour of solutions of wave equations for a  broad class of potentials.  Their results in the global AdS setting imply that there are no power law tails for AdS black holes \cite{Horowitz:1999jd}.

Therefore, as a start, one might expect that the late time decay is governed by the lowest lying quasinormal mode.\footnote{Reference \cite{Chan:1994rs} studied mass inflation for 2+1 dimensional rotating BTZ black hole taking the late time fall-off to be power law. The issue of correct fall-off behaviour was addressed separately in \cite{Chan:1996yk} by the same authors. } The quasinormal mode spectrum for charged black holes in AdS has been explored by many authors \cite{Wang:2000gsa, Wang:2000dt, Berti:2003ud, Wang:2004bv}.  In the large angular momentum limit quasinormal modes in asymptotically globally  AdS spacetimes are  long lived \cite{Festuccia:2008zx, Berti:2009wx, Dias:2012tq}. 
This behaviour is essentially due to the fact that the high angular
momentum parts of a perturbation 
have to tunnel through a large angular momentum barrier.  Holzegel and Smulevici \cite{Holzegel:2011uu} have shown that the decay of a generic scalar perturbation in Kerr-AdS is logarithmic. They argued that this slow decay is due to ``a stable trapping phenomenon'' for Kerr-AdS spacetime, which in turn is a consequence of the reflecting boundary conditions at infinity.  The spectral weight of the perturbation also shifts to higher and higher angular momentum modes as time passes.  These effects  certainly have implications for the internal structure of AdS black holes. A detailed study of all such effects is beyond the scope of the present work. 

Motivated by this, in this section we consider  $(i)$ the exponential fall-off and  $(ii)$ the logarithmic fall-off for the late time behaviour of the mass function $\delta m_1 (v_1)$. We analyse the Ori model for these cases.

\subsection{Exponential fall-off}
To get started, we first assume an exponential decay behaviour of the mass function. Let us take the asymptotic form of $\delta m_1(v_1)$ to be,
\be
\delta m_1 (v_1) \propto e^{-\omega_I v_1}
\ee
where $\omega_I$ is the imaginary part of the lowest quasinormal mode and we take it to be of the form,
\be 
\omega_I=\a \kappa_+,
\ee
where $\a$ is some dimensionless constant, and $\kappa_+$ is the surface gravity of the outer event horizon of the final black hole. Since quasinormal modes at large angular momentum are long lived \cite{Festuccia:2008zx, Berti:2009wx, Dias:2012tq}, $\alpha$ can be as small as we want.

Equation (\ref{linr}), being a first order linear ordinary differential equation,  can be straightforwardly solved using the method of integrating factor. The solution is,
\be
\delta R(v_1)= e^{-\k_- v_1}\left[\int^{v_1} \fr{\delta m_1(\xi)}{r_-^{d-3}} e^{\k_- \xi} d\xi +B \right], \la{If}
\ee
which yields the following expression for $\delta R(v_1)$,
\be
\delta R(v_1)=\fr{A}{(\k_- -\a \kappa_+)} e^{-\a \kappa_+ v_1}+ B e^{-\k_ -v_1},
\ee
where $A, B$ are two constants. We have absorbed certain dimension dependent factors in these constants as well. From equations \eqref{minf2} and \eqref{lamb} we get in the $v_1 \to \infty$ limit,
\be
\Delta m_2(v_1) = - Z_2 \frac{d R(v_1)}{dv_1} \frac{dv_1}{d\l} \propto e^{(\k_- -\a \kappa_+)v_1},
\ee
assuming $\k_- -\a \kappa_+ > 0.$ We see that $\Delta m_2(v_1)$ diverges as $v_1 \to \infty$.
Expressing $v_2$ in terms of $v_1$, 
$v_2\sim -e^{-\k_- v_1}$,
we see that the mass function behaves as,
\be
m_2(v_2) \approx  \Delta m_2(v_2) \propto |-v_2|^{-\fr{ \k_- - \a \kappa_+}{\k_-}}
\la{miexp}.
\ee
For $\k_- - \a \kappa_+ > 0$, the mass function increases without bound as $v_2 \to 0$. The divergence is a power law divergence. This indicates mass inflation.

Since region I and II together cannot be described by a charged Vaidya solution, we introduce double null coordinates $(U, V)$ and determine the metric functions in the mass inflation region. This analysis is similar to the corresponding analysis for the  asymptotically flat case \cite{Ori}. For completeness we present it in appendix \ref{app1}. The metric in these coordinates take the form
\be
ds^2=-2e^{2 \s(U,V)} dU dV + r^2(U,V) d\Omega_{d-2}^2.
\la{DN}
\ee
The coordinate system is set up such  that  the shell $S$ is at $U=0$, and on the shell $V$ is same as the affine parameter $\l$, cf.~\ref{figure1}. Thus, $R(\l):=r(U=0, V=\l)$. Recall that in region II near the shell, the coordinate $v_2$ is also proportional to $\l$, cf.~\eqref{v2_lambda}, therefore 
\be v_2 \propto V.
\ee

In order to express the radial variable $r (U,V)$ as a function of $U, V$ we make use of  equation \eqref{nullc} close to $U=0, V\approx 0$,
\be
 \fr{\partial r}{\partial V}  \propto \fr{\partial r}{\partial v_2} \cong -\fr{m_2(v_2)}{r} \propto - \frac{1}{r} |-V|^{-\fr{ \k_- - \a \kappa_+}{\k_-}}. \label{r_differential2}
\ee
This equation can be integrated to yield,
\be 
r^2(U,V)\cong(r_--U)^2 + 2 c_1  |-V|^{\fr{\a\k_+}{\k_-}} \la{rsq_exp2},
\ee
where $c_1$ is a constant, and we have used the freedom of choosing the coordinate $U$ appropriately (see also discussion around equation \eqref{rsq} in appendix \ref{app1}).  As  $\a\k_+/\k_-$ is positive we can Taylor expand equation (\ref{rsq_exp2}) assuming $U$ and $|-V|^{\fr{\a\k_+}{\k_-}}$ both to be small. Then the function $r$  is given by
\be
r(U,V)=r_- -U\,+\,\fr {c_1}{r_-} |-V|^{\fr{\a\k_+}{\k_-}} + \ldots  \la{rexp2}.
\ee
From equation \eqref{rexp2} we see that for fixed $V$, as $U$ increases the function $r(U,V)$ decreases. This corresponds to the fact that the null generators along constant $V$ lines converge after passing of the outgoing shell $S$.

To obtain the $U, V$ dependence of the function $\s(U,V)$ we use the transformation relations between the metrics (\ref{vaidya}) and (\ref{DN}). This gives
\be 
e^{2\s(U,V)}=-\fr{\partial r}{\partial U}\fr{\partial v_2}{\partial V}.
\la{sigma_reltn}\ee
Taking the logarithm of the above equation and expanding \eqref{rsq_exp2} in Taylor series we get  (see also the corresponding discussion in appendix \ref{app1}),
\be
\sigma(U,V) = \sigma_0 +  c_2 U |-V|^{\fr{\a\k_+}{\k_-}} + \ldots, \label{sigma_exp}
\ee
where  $\sigma_0$ and $c_2$ are constants and we have used the fact that at $U=0$, $V$ is taken to be an affine parameter along the shell.

The mass inflation singularity is a null curvature singularity. Certain curvature components and tidal forces in the freely-falling frame associated to  an infalling observer  diverge as the inner horizon is approached; see appendix \ref{app2} for further details. Tidal distortions can be estimated using the geodesic deviation equation. We find that  tidal forces, i.e., the second time derivative of  tidal distortions, grow as
\be
|-V|^{-2}|-V|^{\frac{\a\k_+}{\k_-}}.
\ee
Integrating this expression twice, we see that  tidal distortions remain finite as $V$ approaches zero.

\subsection{Logarithmic fall-off}

In this sub section,  motivated by the Holzegel and Smulevici \cite{Holzegel:2011uu} analysis,  we consider a  simple minded model where the  asymptotic form of the mass function $\delta m_1(v_1)$ falls-off as $(\log{v_1})^{-2}$,
\be
\delta m_1(v_1) \propto (\log{v_1})^{-2}. \label{log_fall_off}
\ee
 Admittedly, this model misses many features of the Holzegel and Smulevici analysis, but it captures the key point that the decay is logarithmic. With these assumptions we analyse the nature of the mass inflation singularity in the following.

 To determine $R(v_1)$ as a function of $v_1$, as before we Taylor expand $\M(R(v_1))$ in equation (\ref{Req}) around $r=r_-$. We take 
\be
R(v_1)=r_- + \delta R(v_1),
\ee 
to get,
\be
\fr{d}{d v_1} \delta R(v_1) \approx {1\over r
_-^{d-3}}\left[-\k_- r_-^{d-3}\delta R(v_1) + \delta m_1(v_1)\right]. 
\ee
With \eqref{log_fall_off} as the fall-off for the mass function, to leading order in the inverse powers of $v_1$, we have
\be
\delta R(v_1) \approx \fr{\delta m_1(v_1)}{\k_- r_-^{d-3}} \propto  \fr{1}{\k_- r_-^{d-3}  (\log{v_1})^{2}}
\label{linr3}.
\ee

This relation allows us to deduce an expression for mass of the shell, cf.~(\ref{minf2}). Using equation (\ref{lamb}),
\be
\Delta m(v_1)\approx -Z_2\fr{d R(v_1)}{d v_1}\dot{v}_1(\l) \propto \fr{\exp{(\k_- v_1)}}{\k_-^2 r_-^{d-3} v_1 (\log{v_1})^{3}}.
\la{AFmiv1}
\ee
Equivalently, in terms of the affine parameter $\l$ along $S$, 
\be
\Delta m(\l)\propto |\l|^{-1}|(-\log|\l|)|^{-1} |(\log{| \log \l|})|^{-3}.
\ee
Using \eqref{v2_lambda}, we see that  the mass function to leading order takes the following form in region II, 
\be
m_2(v_2)\cong \Delta m(v_2)\approx |v_2|^{-1}|(-\log|v_2|)|^{-1} |(\log{| \log v_2|})|^{-3}. \label{mi_log}
\ee
Expression (\ref{mi_log}) clearly shows the divergent behaviour of the mass parameter in  region II as the inner horizon is approached. This is mass inflation. 

In this case also we can easily determine a regular set of coordinates in the mass inflation region. Doing an analysis similar to what is done in the previous sub section we find,
\be 
r^2(U,V) = (r_-- U)^2 +  2 d_1  \fr{1}{(\log|\log|V||)^2}, \la{rsq_log}
\ee
where $d_1$ is another constant. As $(\log|\log|V||)^{-2}$ is small near $V \approx 0$ we can perform a double Taylor expansion of (\ref{rsq_log}) to obtain
\be
r(U,V)=r_- - U + \fr{d_1}{r_-}(\log|\log|V||)^{-2} + \ldots.
\ee
Similarly, $\s(U,V)$ can be determined using relation (\ref{sigma_reltn}) and expanding (\ref{rsq_log}). This gives,
\be
\s(U,V)=\s_0 + d_2 U (\log|\log|V||)^{-2} + \ldots, 
\ee
where $d_2$ is yet another constant. 

This completes the determination  of a set of regular coordinates in the mass inflation region. In this case also the metric functions remain finite, but certain curvature components and tidal forces in the freely-falling frame associated to  an infalling observer  diverge as the inner horizon is approached; see appendix \ref{app2} for further details. As before, tidal distortions can be estimated using the geodesic deviation equation. We find that  tidal forces grow as
\be
 |-V|^{-2}\times |(-\log|-V|)|^{-1}\times |(\log\left[|-\log|-V|)\right]|^{-3}.
 \ee
Integrating this expression twice, we see that  near $V=0$ tidal distortions  remain finite.

For the logarithmic fall-off case the mass inflation is stronger than the power law case in the sense that the mass function inflates faster as the inner horizon is approached. This is evident from the expressions for $m_2(v_2)$, cf.~\eqref{mi_log} and \eqref{mi}. In the exponential case on the other hand, the growth rate depends on the numerical value of $\alpha$. In the limit when $\alpha$ is zero, the growth rate of the mass function is fastest, cf.~\eqref{miexp}.

\section{Conclusions}
\label{sec:conclusions}
In this paper we have initiated a systematic study of classical internal structure of black holes in AdS spacetime. After reviewing the set-up of the Ori model, we implemented it in the global AdS context.  The construction proceeds by gluing two charged Vaidya spacetimes across an outgoing null shell. The first key element in our analysis is the matching of the two Vaidya coordinates $v_1$ and $v_2$ of the two regions across the shell. The second key element -- required as an input -- is the fall-off behaviour of the mass-function in the external Vaidya spacetime. In this paper we analysed the exponential as well as  the logarithmic fall-off functions. In both cases we found that the mass inflation phenomenon happens, and that the mass inflation singularity is a weak singularity. In fact, we believe that as far as the Ori model is concerned these features are more or less universal. 

Since the mass inflation singularity is weak, it is very intriguing that classical continuation beyond the singularity is not excluded \cite{Ori, Dafermos:2004jp}.  We hope that our study above will inspire further more detailed study of these questions. Numerical simulations and mathematical analysis can help us to understand the mass inflation singularity better. 
In this work we have not analysed the mass inflation singularity for BTZ black holes or for planar AdS black holes. In these cases the fall-off behaviour of the mass function in the external Vaidya spacetime will be different compared to the cases studied above. It will be interesting to analyse these cases and relate them to AdS/CFT discussions. We hope to address some of these questions in our future work.

\subsection*{Acknowledgements}
This project grew out of discussions between AV and Donald Marolf in 2008. AV is grateful to Donald Marolf for suggestions. We are especially grateful to Amos Ori for discussions and explanations on his work. This work was presented at the XIth Field Theoretic Aspects of Gravity (FTAG) meeting in the S.~N.~Bose Center Kolkata and at a seminar in HRI Allahabad. We thank the audience for their useful feedback. SS is supported by the Department of
Science and Technology, Government of India under the Fast Track Scheme for Young Scientists (YSS/2015/001346). 

\appendix
\section{Ori model: further details}
\label{app1}

The considerations of section \ref{sec:ori_set_up} above are quite general. In this appendix we use that set-up to review the mass inflation phenomenon for asymptotically flat black holes and provide some details on the calculations from Ori's paper \cite{Ori}. To obtain the asymptotically flat limit of relevant equations from section \ref{sec:ori_set_up}  we simply take  $l \rightarrow \infty$.

For asymptotically flat boundary conditions Price's law tells the late time behaviour of $\delta m_1(v_1)$ to be \cite{Price:1971fb, Gundlach:1993tp},
\be
\delta m_1(v_1) \propto \frac{1}{v_1^{p-1}},
\ee
with $p \ge 12$ \cite{Poisson:1990eh}.
Therefore, to leading order in the inverse powers of $v_1$, equation \eqref{linr} is solved to be 
\be
\delta R(v_1) \approx \fr{\delta m_1(v_1)}{\k_- r_-^{d-3}}\label{linr2}.
\ee

This relation allows us to deduce an expression for mass of the shell, cf.~(\ref{minf2}). Using equation (\ref{lamb}),
\be
\Delta m(v_1)\approx -Z_2\fr{d R(v_1)}{d v_1}\dot{v}_1(\l) \propto Z_2 \fr{(p-1) \exp{(\k_- v_1)}}{\k_-^2 r_-^{d-3} v_1^{p}}.
\la{AFmiv1}
\ee
Equivalently, in terms of the affine parameter $\l$ along $S$, 
\be
\Delta m(\l)\propto |\l|^{-1}(-\log|\l|)^{-p}.
\ee
In terms of $v_2$, the mass function to leading order takes the form, 
\be
m_2(v_2)\cong \Delta m(v_2)\approx |v_2|^{-1} (-\log|v_2|)^{-p}.\la{mi}
\ee
Expression (\ref{mi}) clearly shows the divergent behaviour of the mass parameter in  region II as the inner horizon is approached.

As the overall geometry cannot be described by a charged Vaidya spacetime, it is convenient to introduce double null coordinates and write the full spacetime metric as
\be
ds^2=-2e^{2 \s(U,V)} dU dV + r^2(U,V) d\Omega_{d-2}^2,
\la{dn}
\ee
where the metric functions  $\s(U,V)$ and $r(U,V)$ are to be determined.  We require that the shell $S$ is at $U=0$, moreover that on the shell $V$ is same as the affine parameter $\l$. Then, $R(\l):=r(U=0, V=\l)$. In region II near the shell, the coordinate $v_2$ is also proportional to $\l$, cf.~\eqref{v2_lambda}, therefore \be v_2 \propto V.\ee

Now we turn to determine the solution in the mass inflation region. For this we have to express $r$ and $\s$ as functions of $U$ and $V$. We can find the $V$ dependence of $r(U,V)$ in the mass inflation region (the vicinity of $U=0, V=0$) by using the relation \eqref{nullc},
\be
\fr{\partial r}{\partial v_2} \propto \fr{\partial r}{\partial V} \cong -\fr{m_2(v_2)}{r} \approx -\fr{1}{r|V|}\fr{1}{(-\log|V|)^p}, \label{r_differential}
\ee
where we have used the fact that near $V=0$ the mass function \eqref{mi} diverges and the radial variable $r$ remains finite.
Equation \eqref{r_differential} can be integrated to give
\be
r^2(U,V) = 2 k_1  (-\log |V|)^{1-p} + H(U).
\la{rsq}
\ee 
The arbitrary function $H(U)$ reflects the freedom in choosing the $U$ coordinate, and $k_1$ is a proportionality constant.
We choose
\be
H(U) = (r_- - U)^2,
\ee 
therefore, 
\be
r^2 = (r_- - U)^2 + 2 k_1  (-\log |V|)^{1-p}. \label{req_not_expanded}
\ee

Now we assume that both $U$ and $(-\log |V|)^{1-p}$ are small --- and the smallness is of the same ``level''. Then Taylor expansion gives,
\be
r(U,V) = r_- - U +  \frac{k_1}{r_-}  (-\log |V|)^{1-p} + \ldots \label{req}.
\ee
The above expression is same as equation (16) of Ori's paper \cite{Ori}.

To find $\s(U,V)$ we use the coordinate transformations from $(r,v_2)$ set of coordinates to $(U,V)$ coordinates. This gives,
\be 
e^{2\s(U,V)}=-\fr{\partial r}{\partial U}\fr{\partial v_2}{\partial V}.
\ee
Taking the logarithm of the above equation we have,
\be
2 \s(U,V) = \log \left[-\frac{\partial r}{\partial U} \frac{dv_2}{dV}\right].
\ee
Using \eqref{req_not_expanded} and Taylor expanding to third order we find
\be
\sigma(U,V) = \sigma_0 +k_2   U (-\log |V|)^{1-p} + \ldots, \label{sigma}
\ee
where  $\sigma_0$ and $k_2$ are constants and we have used the fact that at $U=0$, $V$ is taken to be an affine parameter along the shell. 
Together with equation \eqref{req} this completes the determination of the geometry in the mass inflation region. Equations \eqref{req} and \eqref{sigma} were stated in Ori's paper without any details. The above details fill-in this gap in the literature\,\footnote{We are grateful to A.~Ori for a detailed discussion on this point.}.

\section{Strength of the mass inflation singularity}
\label{app2}
In this appendix we address the question of the strength of the mass inflation singularity. Let us consider a freely falling observer  with velocity vector $u^\alpha$  in background \eqref{DN} approaching the inner horizon $V=0$. The freely-falling frame associated to the observer is 
\bea
e^\alpha_{(0)} &=& u^\alpha = \frac{e^{-\sigma}}{\sqrt{2}} \left(1/u^V,u^V,0,0, \ldots\right), \\
e^\alpha_{(1)} &=&  \frac{e^{-\sigma}}{\sqrt{2}} \left(-1/u^V,u^V,0,0,\ldots\right), \\
e^\alpha_{(2)} &=&  \left(0,0 ,\frac{1}{r},0,\ldots\right), \\
e^\alpha_{(3)} &=&  \left(0,0,0,\frac{1}{r \sin \theta}, \ldots\right), \\
e^\alpha_{(4)} &=& \ldots, \nn \\
  & \vdots & \nn
\eea
with $u^\alpha = \frac{dx^\a}{d\tau}$, where $\tau$ is the proper time.  This frame is parallel transported along the timelike geodesic of the observer.

If the deviation vector connecting a point in a test body to its centre of mass is denoted by $\epsilon^\alpha$, then the geodesic deviation equation tells us
\be
\frac{D^2 \epsilon^\alpha}{d\tau^2} = - R^\alpha{}_{\beta \gamma \delta} u^\beta  \epsilon^\gamma u^\delta.
\ee
We consider the body to be spacelike, so we project $\epsilon^\alpha$ along the spacelike vielbeins, and define
\be
x^i = e^{(i)}_\alpha  \epsilon^\alpha, \qquad i = 1, 2, \ldots, d-1.
\ee
Since  $ e^{(i)}_\alpha $  are parallel transported along the geodesic, the geodesic deviation equation implies
\be
\frac{D^2 x^i}{d\tau^2}  + K^{i}{}_j x^j= 0, \label{geodesic_deviation}
\ee
where
\be
K_{ij} =  R_{\alpha \beta \gamma \delta} e^\alpha_{(i)}  u^\beta  e^\gamma_{(j)} u^\delta.
\ee
These are  equations (5) and (6) of reference \cite{Hiscock}.  A direct calculation allows us to find various components of $K_{ij}$. We find
\bea
K_{11} &=& - 2 e^{-2 \s} \s_{, UV}, \\
K_{22} &=& \frac{e^{-2 \s}}{r} \left[ r_{,V} \s_{,V} (u^V)^2 - r_{, UV} + \frac{r_{,U} \s_{,U}}{(u^V)^2}  - \frac{1}{2} \left( r_{,VV} (u^V)^2 + \frac{r_{,UU}}{(u^V)^2 }  \right)\right], \\
K_{33} &=& K_{44} = \ldots = K_{22}.
\eea

In order to proceed we need to find the relation of $u^V$ to proper time $\tau$. The geodesic equation gives
\be
\frac{du^V}{d\tau} + \Gamma^V_{VV} \left(u^V\right)^2 =0, \label{geodesicV}
\ee
where 
\be
\Gamma^V_{VV} = 2 \sigma_{,V}. 
\ee
A solution to this system (near $V=0$ in all cases of interest) to leading order is
$V \approx \tau$. 
With this one finds that in the limit $V \to 0$, the leading divergence in $K_{22}$ is 
$V^{-2} \left|\log|-V| \right|^{-p}$ for the power law fall-off \cite{Ori}.
For the exponential fall-off case we get,
\be
K_{22}\propto |-V|^{-2}|-V|^{\frac{\a\k_+}{\k_-}},
\ee
and similarly, for the logarithmic fall-off we get 
\be
K_{22}\propto |-V|^{-2}\times |(-\log|-V|)|^{-1}\times |(\log\left[|-\log|-V|)\right]|^{-3}.
\ee
In all cases, two integration of \eqref{geodesic_deviation} gives a $V$ dependence that goes to zero as $V$ goes to zero. Therefore, in all cases,  tidal distortions are finite -- the mass inflation singularity is a weak singularity; continuation beyond it is classically not forbidden \cite{Ori}. Naturally, this is related to the fact that the metric functions $\sigma$ and $r$ as well as  the metric determinant all remain finite at $V=0$, but their $V$ derivatives are not necessarily finite.


\begin{thebibliography}{99}
  
  
\bibitem{Penrose:68}
R.~Penrose, in ``Batelle Rencontres,'' eds. C.~De
Witt and J.~Wheeler (W. A. Benjamin, 1968).

\bibitem{Matzner}
R.~A.~Matzner, N.~Zamorano and V.~D.~Sanberg, ``Instability of the
Cauchy horizon of Reissner-Nordstr\"{o}m black holes,'' Phys.\ Rev.\ D
{\bf 19}, 2821 (1979).

\bibitem{Chandra:82}
S.~Chandrasekhar and J.~Hartle, ``On crossing the Cauchy horizon of a
Reissner-Nordstr\"{o}m black-hole,'' Proc.\ Roy.\ Soc.\
Lond.\ {\bf A384}, 301 (1982).


\bibitem{Marolf:2011dj} 
  D.~Marolf and A.~Ori,
``Outgoing gravitational shock-wave at the inner horizon: The late-time limit of black hole interiors,''
  Phys.\ Rev.\ D {\bf 86}, 124026 (2012)
  doi:10.1103/PhysRevD.86.124026
  [arXiv:1109.5139 [gr-qc]].



\bibitem{Brady:1995ni} 
  P.~R.~Brady and J.~D.~Smith,
  ``Black hole singularities: A Numerical approach,''
  Phys.\ Rev.\ Lett.\  {\bf 75}, 1256 (1995)
  doi:10.1103/PhysRevLett.75.1256
  [gr-qc/9506067].
  
\bibitem{Burko:1997zy} 
  L.~M.~Burko,
  ``Structure of the black hole's Cauchy horizon singularity,''
  Phys.\ Rev.\ Lett.\  {\bf 79}, 4958 (1997)
  doi:10.1103/PhysRevLett.79.4958
  [gr-qc/9710112].
  
\bibitem{Hod:1998gy} 
  S.~Hod and T.~Piran,
  ``Mass inflation in dynamical gravitational collapse of a charged scalar field,''
  Phys.\ Rev.\ Lett.\  {\bf 81}, 1554 (1998)
  doi:10.1103/PhysRevLett.81.1554
  [gr-qc/9803004].
      

\bibitem{Dafermos:2003wr} 
  M.~Dafermos,
  ``The Interior of charged black holes and the problem of uniqueness in general relativity,''
  Commun.\ Pure Appl.\ Math.\  {\bf 58}, 0445 (2005)
  [gr-qc/0307013].
  

\bibitem{Dafermos:2003yw} 
  M.~Dafermos and I.~Rodnianski,
  ``A Proof of Price's law for the collapse of a selfgravitating scalar field,''
  Invent.\ Math.\  {\bf 162}, 381 (2005)
  doi:10.1007/s00222-005-0450-3
  [gr-qc/0309115].
  
\bibitem{Dafermos:2004jp} 
  M.~Dafermos,
  ``Price's law, mass inflation, and strong cosmic censorship,''
in  Proceedings of 7th Hungarian Relativity Workshop (RW 2003): Sarospatak, Hungary, August 10-15, 2003
I. Racz (ed.),  gr-qc/0401121.


\bibitem{Poisson:1990eh} 
  E.~Poisson and W.~Israel,
  ``Internal structure of black holes,''
  Phys.\ Rev.\ D {\bf 41}, 1796 (1990).
  doi:10.1103/PhysRevD.41.1796


\bibitem{Price:1971fb} 
  R.~H.~Price,
 ``Nonspherical perturbations of relativistic gravitational collapse. 1. Scalar and gravitational perturbations,''
  Phys.\ Rev.\ D {\bf 5}, 2419 (1972).
  doi:10.1103/PhysRevD.5.2419
  
  
\bibitem{Gundlach:1993tp} 
  C.~Gundlach, R.~H.~Price and J.~Pullin,
  ``Late time behavior of stellar collapse and explosions: 1. Linearized perturbations,''
  Phys.\ Rev.\ D {\bf 49}, 883 (1994)
  doi:10.1103/PhysRevD.49.883
  [gr-qc/9307009].
  

\bibitem{Hiscock} 
William A. Hiscock,  ``Evolution of the interior of a charged black hole,'' Physics Letters A
Volume 83, Issue 3, 18 May 1981, Pages 110-112.

 
\bibitem{Ori} 
  A.~Ori,
  ``Inner structure of a charged black hole: An exact mass-inflation solution,''
  Phys.\ Rev.\ Lett.\  {\bf 67}, 789 (1991).
  doi:10.1103/PhysRevLett.67.789
  
  
\bibitem{Horowitz:1999jd} 
  G.~T.~Horowitz and V.~E.~Hubeny,
  ``Quasinormal modes of AdS black holes and the approach to thermal equilibrium,''
  Phys.\ Rev.\ D {\bf 62}, 024027 (2000)
  doi:10.1103/PhysRevD.62.024027
  [hep-th/9909056].
  
    
  
\bibitem{Chan:1994rs} 
  J.~S.~F.~Chan, K.~C.~K.~Chan and R.~B.~Mann,
  ``Interior structure of a charged spinning black hole in (2+1)-dimensions,''
  Phys.\ Rev.\ D {\bf 54}, 1535 (1996)
  doi:10.1103/PhysRevD.54.1535
  [gr-qc/9406049].
  
  
  
\bibitem{Chan:1996yk} 
  J.~S.~F.~Chan and R.~B.~Mann,
  ``Scalar wave falloff in asymptotically anti-de Sitter backgrounds,''
  Phys.\ Rev.\ D {\bf 55}, 7546 (1997)
  doi:10.1103/PhysRevD.55.7546
  [gr-qc/9612026].

  
\bibitem{Festuccia:2008zx} 
  G.~Festuccia and H.~Liu,
  ``A Bohr-Sommerfeld quantization formula for quasinormal frequencies of AdS black holes,''
  Adv.\ Sci.\ Lett.\  {\bf 2}, 221 (2009)
  doi:10.1166/asl.2009.1029
  [arXiv:0811.1033 [gr-qc]].
  
\bibitem{Berti:2009wx} 
  E.~Berti, V.~Cardoso and P.~Pani,
  ``Breit-Wigner resonances and the quasinormal modes of anti-de Sitter black holes,''
  Phys.\ Rev.\ D {\bf 79}, 101501 (2009)
  doi:10.1103/PhysRevD.79.101501
  [arXiv:0903.5311 [gr-qc]].
  
\bibitem{Dias:2012tq} 
  O.~J.~C.~Dias, G.~T.~Horowitz, D.~Marolf and J.~E.~Santos,
  ``On the Nonlinear Stability of Asymptotically Anti-de Sitter Solutions,''
  Class.\ Quant.\ Grav.\  {\bf 29}, 235019 (2012)
  doi:10.1088/0264-9381/29/23/235019
  [arXiv:1208.5772 [gr-qc]].

\bibitem{Holzegel:2011uu} 
  G.~Holzegel and J.~Smulevici,
  ``Decay properties of Klein-Gordon fields on Kerr-AdS spacetimes,''
  Commun.\ Pure Appl.\ Math.\  {\bf 66}, 1751 (2013)
  doi:10.1002/cpa.21470
  [arXiv:1110.6794 [gr-qc]].

 
\bibitem{Brecher:2004gn} 
  D.~Brecher, J.~He and M.~Rozali,
 ``On charged black holes in anti-de Sitter space,''
  JHEP {\bf 0504}, 004 (2005)
  doi:10.1088/1126-6708/2005/04/004
  [hep-th/0410214].

   
\bibitem{Barrabes:1991ng} 
  C.~Barrabes and W.~Israel,
 ``Thin shells in general relativity and cosmology: The Lightlike limit,''
  Phys.\ Rev.\ D {\bf 43}, 1129 (1991).
  doi:10.1103/PhysRevD.43.1129
   

\bibitem{Poisson} 
E. Poisson, {\it A Relativist's Toolkit: The Mathematics of Black-Hole Mechanics}, 2004 (Cambridge University Press).


\bibitem{Ching:1995tj} 
  E.~S.~C.~Ching, P.~T.~Leung, W.~M.~Suen and K.~Young,
 ``Wave propagation in gravitational systems: Late time behavior,''
  Phys.\ Rev.\ D {\bf 52}, 2118 (1995)
  doi:10.1103/PhysRevD.52.2118
  [gr-qc/9507035].
  
\bibitem{Wang:2000gsa} 
  B.~Wang, C.~Y.~Lin and E.~Abdalla,
  ``Quasinormal modes of Reissner-Nordstrom anti-de Sitter black holes,''
  Phys.\ Lett.\ B {\bf 481}, 79 (2000)
  doi:10.1016/S0370-2693(00)00409-3
  [hep-th/0003295].
  
\bibitem{Wang:2000dt} 
  B.~Wang, C.~Molina and E.~Abdalla,
  ``Evolving of a massless scalar field in Reissner-Nordstrom Anti-de Sitter space-times,''
  Phys.\ Rev.\ D {\bf 63}, 084001 (2001)
  doi:10.1103/PhysRevD.63.084001
  [hep-th/0005143].
  
  
\bibitem{Berti:2003ud} 
  E.~Berti and K.~D.~Kokkotas,
  ``Quasinormal modes of Reissner-Nordstrom-anti-de Sitter black holes: Scalar, electromagnetic and gravitational perturbations,''
  Phys.\ Rev.\ D {\bf 67}, 064020 (2003)
  doi:10.1103/PhysRevD.67.064020
  [gr-qc/0301052].
  
\bibitem{Wang:2004bv} 
  B.~Wang, C.~Y.~Lin and C.~Molina,
  ``Quasinormal behavior of massless scalar field perturbation in Reissner-Nordstrom anti-de Sitter spacetimes,''
  Phys.\ Rev.\ D {\bf 70}, 064025 (2004)
  doi:10.1103/PhysRevD.70.064025
  [hep-th/0407024].
  

  \end{thebibliography}
\end{document}